\RequirePackage{fix-cm}
\documentclass[]{svjour3}                     

\smartqed  

\usepackage{graphicx}
\usepackage{amsmath}
\usepackage{mathrsfs}
\usepackage{cite}
\usepackage{verbatim}
\usepackage{marvosym}
\usepackage{latexsym}
\usepackage{lineno,hyperref}
\usepackage{amssymb}
\usepackage[noend]{algpseudocode}
\usepackage{algorithmicx,algorithm}
\usepackage{subfigure}
\usepackage{graphicx}  
\usepackage{float}  
\usepackage{subfigure}  
\begin{document}

\title{Quantum adversarial metric learning model based on triplet loss function  
}
\subtitle{}

\titlerunning{Quantum adversarial metric learning model based on triplet loss function}        

\author{Yan-Yan Hou    $^{\textbf{1,2}}$    \and
        Jian Li        $^{\textbf{3}}$        \and
        Xiu-Bo Chen    $^{\textbf{4,5}}$      \and
        Chong-Qiang Ye  $^{\textbf{1}}$        \and
}

\institute{
\Letter Jian Li\\
Lijian@bupt.edu.cn\\
$^{\textbf{1}}$School of Artificial Intelligence, Beijing University of Posts and Telecommunications, Beijing 100876, China.\\
$^{\textbf{2}}$College of Information Science and Engineering, ZaoZhuang University, ZaoZhuang Shandong 277160, China.\\
$^{\textbf{3}}$School of Cyberspace Security Security, Beijing University of Posts Telecommunications, Beijing 100876, China.\\
$^{\textbf{4}}$Information Security Center, State Key Laboratory of Networking and Switching Technology, Beijing University of Post and Telecommunications, Beijing 100876, China.\\
$^{\textbf{5}}$GuiZhou University, Guizhou Provincial Key Laboratory of Public Big Data, Guizhou Guiyang, 550025, China.\\
}

\date{Received: date / Accepted: date}
\maketitle
\begin{abstract}

Metric learning plays an essential role in image analysis and classification, and it has attracted more and more attention. In this paper, we propose a quantum adversarial metric learning (QAML) model based on the triplet loss function, where samples are embedded into the high-dimensional Hilbert space and the optimal metric is obtained by minimizing the triplet loss function. The QAML model employs entanglement and interference to build superposition states for triplet samples so that only one parameterized quantum circuit is needed to calculate sample distances, which reduces the demand for quantum resources. Considering the QAML model is fragile to adversarial attacks, an adversarial sample generation strategy is designed based on the quantum gradient ascent method, effectively improving the robustness against the functional adversarial attack. Simulation results show that the QAML model can effectively distinguish samples of MNIST and Iris datasets and has higher $\epsilon$-robustness accuracy over the general quantum metric learning. The QAML model is a fundamental research problem of machine learning. As a subroutine of classification and clustering tasks, the QAML model opens an avenue for exploring quantum advantages in machine learning.


\end{abstract}

\section{Introduction}
\label{intro}
Machine learning has developed rapidly in recent years and is widely used in artificial intelligence and big data fields. Quantum computing can efficiently process data in exponentially sizeable Hilbert space and is expected to achieve dramatic speedups in solving some classical computational problems. Quantum machine learning, as the interplay between machine learning and quantum physics, brings unprecedented promise to both disciplines. On the one hand, machine learning methods have been extended to quantum world and applied to the data analysis in quantum physics\cite{cong2019quantum}. On the other hand, quantum machine learning exploits quantum properties, such as entanglement and superposition, to revolutionize classical machine learning algorithms and achieves computational advantages over classical algorithms\cite{benedetti2019parameterized}. Metric Learning is the core problem of some machine learning tasks\cite{chen2018adversarial}, such as $k$-nearest neighbor, support vector machines, radial basis function networks, and $k$-means clustering. Its core work is to construct an appropriate distance metric that maximizes the similarities of samples of the same class and minimizes the similarities of samples from different classes. Linear and nonlinear methods can be used to implement metric learning. In general, linear models have a limited number of parameters and are unsuitable for characterizing high-order features of samples. Recently, neural networks have been adopted to establish nonlinear metric learning models, and promising results have been achieved in face recognition and feature matching.

Classical metric learning models usually extract low-dimensional representations of samples, which will lose some details of samples. Quantum states are in high-dimensional Hilbert spaces, and their dimensions grow exponentially with the number of qubits. This quantum enables quantum models to learn high-dimensional representations of samples without explicitly invoking a kernel function. A parameterized quantum circuit is used to map samples in high-dimensional Hilbert space. The optimal metric model is obtained by optimizing the loss function based on Hilbert-Schmidt distances. With the increase of the the dimension, this speed-up advantage will become more and more pronounced, and it is expected to achieve exponential growth in computing speeds.
In recent years, researchers began to study how to adopt quantum methods to implement metric learning. Lloyd\cite{lloyd2020quantum} firstly proposed a quantum metric learning model based on hybrid quantum-classical algorithms. A parameterized quantum circuit is used to map samples in high-dimensional Hilbert space. The optimal metric model is obtained by optimizing the loss function based on Hilbert-Schmidt distances. This model achieves better effects in classification tasks. Nhat\cite{nghiem2020unified} introduced quantum explicit and implicit metric learning approaches from the perspective of whether the target space is known or not. The research establishes the relationship between quantum metric learning and other quantum supervised learning models. The above two algorithms mainly focus on classification tasks. Metric learning is a fundamental problem in machine learning, which can be applied not only to classification but also to clustering, face recognition, and other issues. In our research, we are devoted to constructing a quantum metric learning model that can serve various machine learning tasks.

Angular distance is a vital metric that quantifies the included angle between normalized samples\cite{mao2019metric}. Angular distance focuses on the difference in the direction of samples and is more robust to the variation of local feature\cite{wang2017deep},\cite{duan2018deep}. Considering the similarities between angular distances of classical data and inner products of quantum states, we design a quantum adversarial metric learning (QAML) model based on inner product distances, which is more suitable for image-related tasks. Unlike other quantum metric learning models, the QAML model maps samples from different classes into quantum superposition states and utilizes simple interface circuits to compute metric distances for multiple sample pairs in parallel. Furthermore, quantum systems in high-dimensional Hilbert space have counter-intuitive geometrical properties\cite{liu2020vulnerability}. The QAML model using only natural samples is vulnerable to adversarial attacks, under which some samples are closer to the false class, so the model is easy to make wrong judgements\cite{madry2017towards}. To solve this issue, we construct adversarial samples based on natural samples. The model's robustness is improved by the alternative train of natural and adversarial samples.
Our work has two main contributions:(i) We explore a quantum method to compute the triplet loss function, which utilizes quantum superposition states to calculate sample distances in parallel and reduce the demand for quantum resources. (ii) We design an adversarial samples generation strategy based on the quantum gradient ascent, and the robustness of the QAML model is significantly improved by alternatively training generated adversarial samples and natural samples. Simulation results show that the QAML model separates samples by a larger margin and has better robustness for functional adversarial attacks than general quantum metric learning models.

The paper is organized as follows. Section 2 gives the basic method of the QAML model. Section 3 verifies the performances of the QAML model. Finally, we get a conclusion and discuss the future research directions.

\section{Quantum adversarial metric learning}
\subsection{Preliminary theory}
Triplet loss function is a widely used strategy for metric learning\cite{salakhutdinov2007learning}, commonly used in image retrieval and face recognition. A triplet set ${(x_i^a, x^p_i, x^n_i)}$ consists of three samples from two classes, where anchor sample $x_i^a$ and positive sample $x^p_i$ belong to the same class, and negative sample $x^n_i$ comes from another class. The goal of metric learning based on triplet loss function is to find the optimal embedded representation space, in which positive sample pairs ${(x_i^a, x^p_i)}$ are pulled together and negative sample pairs ${(x_i^a, x^n_i)}$ are pushed away. Fig.1 shows sample space change in the metric learning process. As we can see, samples from different classes become linearly separable through metric learning. Fig.2 shows the schematic of the metric learning model based on triplet loss function. Firstly, the model prepares multiple triplet sets, and one triplet set ${(x_i^a, x^p_i, x^n_i)}$ is sent to convolutional neural networks (CNN), where three CNN with the same structure and parameters are needed. Each CNN acts on one sample of the triplet set to extract its features. The triplet loss function is obtained by computing metric distances for multiple sample pairs of triplet sets. In the learning process, the optimal parameters of CNN are obtained by minimizing the triplet loss function.
\begin{figure*}[t]
\centering
\includegraphics[scale=0.6]{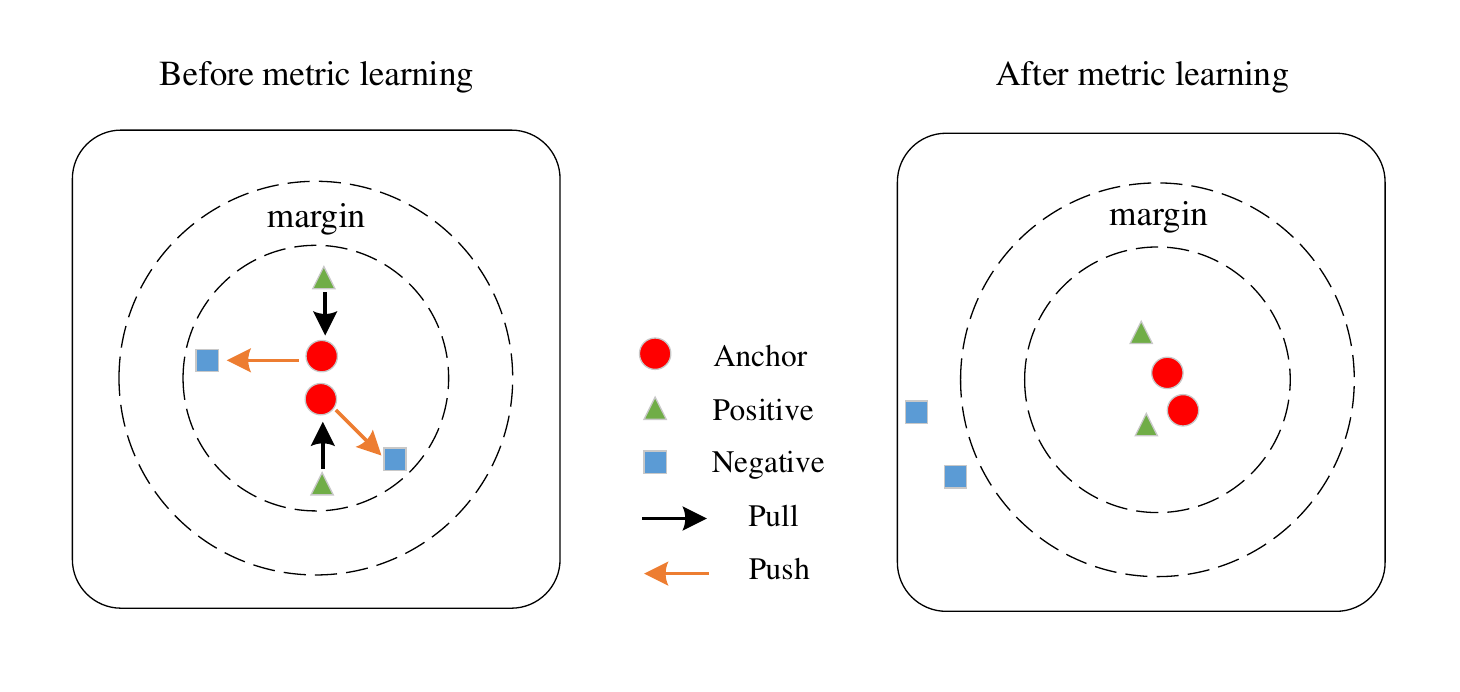}
\vspace{0cm}
\caption{Sample space change in metric learning process. Before metric learning, the distances between negative sample pairs are smaller, and samples from different classes are difficult to separate by linear functions. After metric learning, the distances between negative sample pairs become larger, and a large margin separates samples from different classes. Linear functions can easily separate positive and negative samples.}
\label{figure2}
\vspace{0cm}
\end{figure*}
\begin{figure*}[t]
\centering
\includegraphics[scale=0.6]{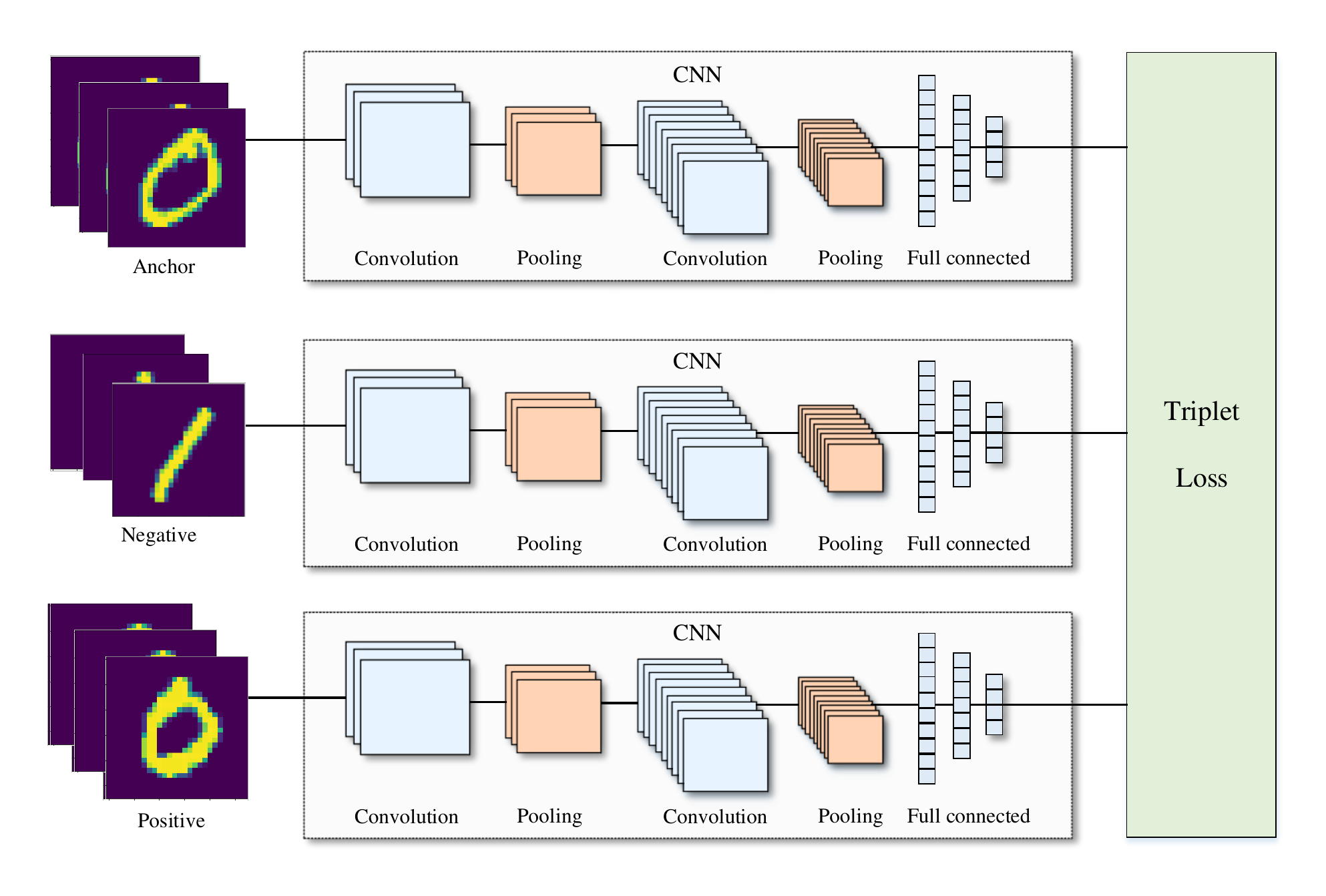}
\vspace{0cm}
\caption{The schematic of the metric learning model based on triplet loss function. A triplet set includes an anchor sample, a positive sample, and a negative sample. The input consists of a batch of triplet sets, and only one triplet set serves as input in each iteration. Three CNN with the same structure and parameters are used to map the triplet set into the embedded representation space. CNN, consisting of multiple convolutions, pooling, and fully connected layers, is responsible for extracting the features of samples. The triplet loss function is further constructed based on the extracted features.}
\label{figure2}
\vspace{0cm}
\end{figure*}
Let one batch samples include $N_1$ triplet sets. The triplet loss function is
\begin{equation}
\label{E41}
\begin{aligned}
\begin{array}{ccc}
L= \frac{1}{N_1}\sum_{i=1}^{N_1}[D(g(x^a_i),g(x^p_i))-D(g(x^a_i),g(x^n_i))+\mu]_{+},
\end{array}
\end{aligned}
\end{equation}
where $g(\cdot)$ represents the function mapping input samples to the embedded representation space, $D(\cdot,\cdot)$ denotes the distance between a sample pair in the embedded representation space, and $[\ \cdot\ ,\ \cdot\ ]_{+}=max(0,\ \cdot\ )$ represents the hinge loss function. The goal of metric learning is to learn a metric that makes the distances between negative sample pairs greater than the distance between the corresponding positive sample pairs and satisfies the specified margin $\mu\in \mathbb{R}^+$\cite{mao2019metric}. In the triplet loss function, $D(g(x^a_i), g(x^p_i))$ penalizes the positive sample pair $(x^a_i, x^p_i)$ that is too far apart, and $D(g(x^a_i), g(x^n_i))$ penalizes the negative sample pair $(x^a_i,x^p_i)$ whose distance is less than the margin $\mu$.

Metric learning can adopt various distance metric methods. Angular distance metric is robust to image illumination and contrast variation \cite{wang2017deep}, which is an efficient way for metric learning tasks. In this method, samples need to be normalized to unit vectors in advance. The distance between a positive sample pair is
\begin{equation}
\label{E21}
\begin{aligned}
\begin{array}{ccc}
D(g(x^a_i), g(x^p_i))=1-\frac{|g(x^a_i)\cdot g(x^p_i)|}{||g(x^a_i)||_2||g(x^p_i)||_2},
\end{array}
\end{aligned}
\end{equation}
where $|\ |$ and $||\ ||_2$ represent $l_1$-norm and $l_2$-norm, respectively, and $\cdot$ denotes the inner product operation for two vectors. The distance between negative sample pairs can be calculated in the same way.

\subsection{Framework of quantum metric learning model}

For most machine learning tasks, it is often challenging to adopt simple linear functions to distinguish samples of different classes. According to kernel theory\cite{blank2020quantum}, samples in high-dimensional feature space have better distinguishability. Classical machine learning algorithms usually adopt kernel methods to map samples to high-dimensional feature space, where the mapped samples can be separated by simple linear functions. Quantum states with $n$-qubits are in $2^n$-dimensional Hilbert space, where quantum systems characterize the nonlinear features of data and efficiently process data through a series of linear unitary operations.

In the QAML model, samples should be firstly mapped into quantum systems by qubit encoding. The Hilbert space after encoding usually does not correspond to the optimal space for separating samples of different classes. To search for the optimal Hilbert space, the QAML model performs parameterized quantum circuits $W(\theta)$ on the encoded states\cite{grant2018hierarchical}. As different variable parameters $\theta$ correspond to different mapping spaces, we can search the optimal space by modifying parameters $\theta=(\theta_1^1,...,\theta_i^j)$. As long as $W(\theta)$ has strong expressivity, we can find the optimal Hilbert space by optimizing the loss function of metric learning\cite{perez2020data,schuld2021effect}. $W(\theta)$ with different structures and layers have different expressivity. The more layers $W(\theta)$ has, the stronger the expressivity, and the easier it is to find the optimal metric space.

The classical metric learning model based on triplet loss function requires three identical CNN to map triplet sets $(x_i^a, x_i^p, x_i^n)$ into the novel Hilbert space. To reduce the demand for quantum resources, we construct a quantum superposition state to represent one triplet set so that a triplet set only needs one $W(\theta)$ to transform it into Hilbert space. The core work of the building loss function is to compute inner products between sample pairs, but $W(\theta)$ and subsequent conjugate operation $W^{\dagger}(\theta)$ counteract each other's effects. To solve this issue, we add a repeated encoding operation after $W(\theta)$. It is worth mentioning that the repeated encoding operation is also conducive to the construction of high-dimensional features of samples.

The QAML model is mathematically represented as the minimization of the loss function with respect to the parameters $\theta$. The triplet loss function consists of metric distances for positive and negative sample pairs, so the kernel work of the QAML model is constructing the metric distances for sample pairs in the transformed Hilbert space. The mapping samples $h(x_i^a)/||h(x_i^a)||_2$ and $h(x_i^p)/||h(x_i^p)||_2$ of Equ.\ref{E21} are replaced by the quantum states of $x_i^a$ and $x_i^p$, then the second term of Equ.\ref{E21} is converted to the inner product between quantum states of the positive sample pair $(x_i^a, x_i^p)$, which can be got by the method of the Hadamard classifier\cite{blank2020quantum}. The triplet loss function can be viewed as the weighted sum of the inner product of sample pairs $(x_i^a, x_i^p)$ and the inner product of sample pairs $(x_i^a,x_i^n)$. With the help of ancilla registers, the triplet set can be prepared in superposition states form. According to the entanglement property of superposition states, the triplet loss function can be implemented with one parameterized quantum circuit. Then, the triplet loss function value is transmitted to a classical optimizer, and parameters are optimized until the optimal metric is obtained. The QAML model constructs adversarial samples according to the gradient of natural samples and trains alternatively natural and adversarial samples to improve the model's robustness against adversarial attacks. The schematic of the QAML model is shown in Fig.3.

\begin{figure*}[t]
\centering
\includegraphics[scale=0.5]{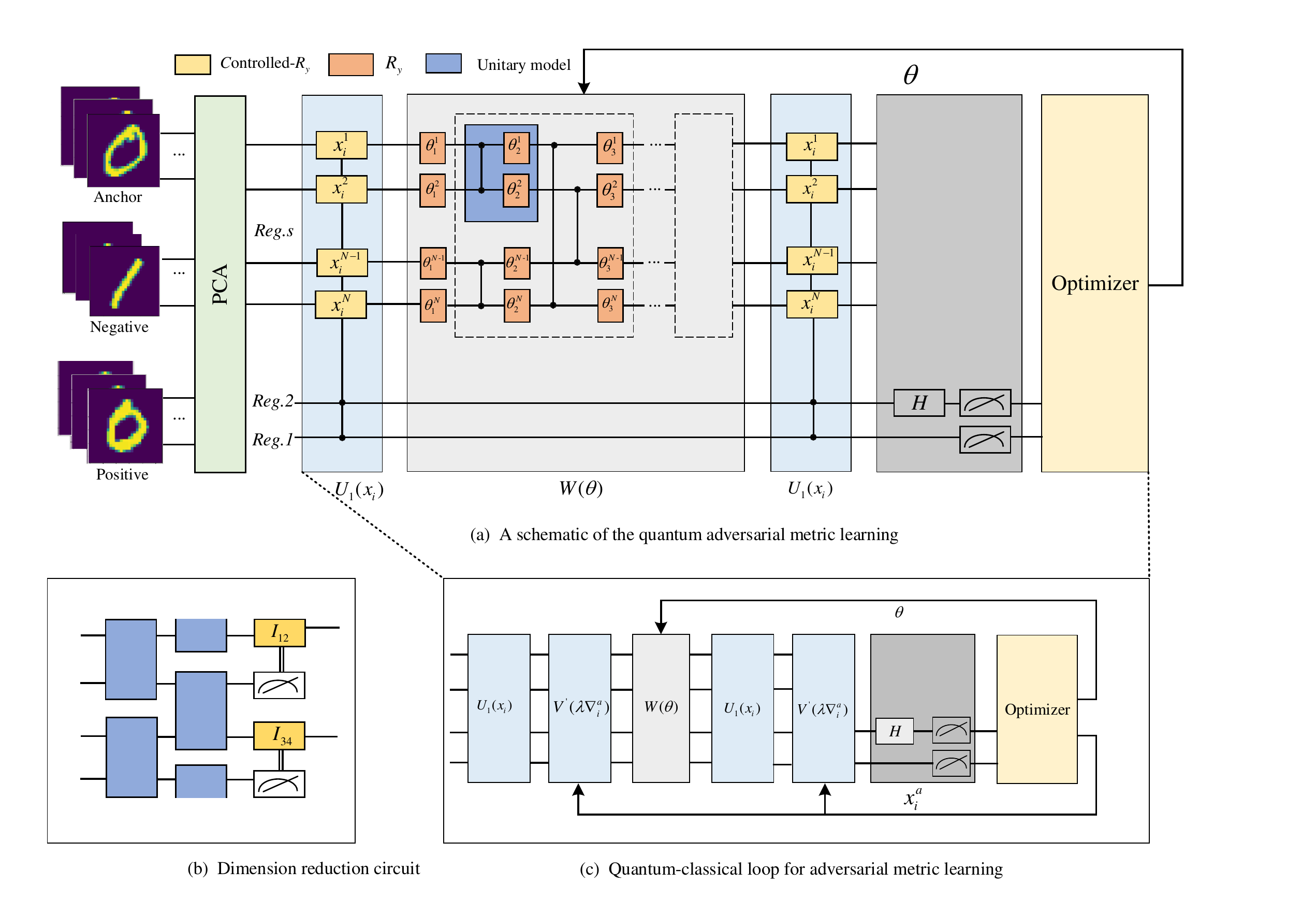}
\vspace{0cm}
\caption{Overview of quantum adversarial metric learning (QAML) model. Panel (a) shows the framework of quantum adversarial metric learning. $Reg.s$ is the sample register that stores triplet sets, and $Reg.1$ and $Reg.2$ are ancilla registers used distinguishing different samples. The model firstly adopts principal component analysis (PCA) to reduce the input dimension. Subsequently, anchor, negative and positive samples are encoded into a quantum superposition state by controlled qubit encoding. The transformation of Hilbert space is implemented by parameterized quantum circuit $W(\theta)$ and the subsequent qubit encoding $U_1(x_i)$. Finally, Hadamard and measurement operations act on ancilla registers to simultaneously compute the inner products for the positive and negative sample pairs, and the triplet loss function is further obtained. In each iteration, the parameters $\theta$ are updated by optimizing the triplet loss function with a classical optimizer. Panel (b) shows the quantum dimension reduction circuit to reduce the number of output qubits. In each module, only one qubit is measured, and the controlled unitary based on its measurement result acts on another qubit. Panel (c) shows another case of the QAML model, where adversarial samples are built and added to the training process. $V^{'}(\lambda\nabla_i^a)$ is the unitary operation based on the gradient of anchor sample $x_i^a$ and acts on the encoded quantum states to produce its adversarial sample. In the QAML model training process, natural and adversarial samples alternatively serve as input.}
\label{figure2}
\vspace{0cm}
\end{figure*}

\subsection{Quantum embedding}
In the QAML model, classical samples are firstly mapped into quantum states by qubit encoding, where each element is encoded as a Pauli rotation angle of one qubit. The number of qubits required by qubit encoding is equivalent to the dimension of the input sample. Still, the dimension of one quantum state grows exponentially with the input dimension, and $N$-dimensional samples will be mapped to $2^{N}$-dimensional Hilbert space. The qubit encoding method cannot use logarithm qubits of the input sample dimension to represent classical samples. However, easy state preparation and low circuit depth make qubit encoding more suitable for implementation on near-term quantum devices.

Samples in practical applications are usually in real space. Applying $R_X$ and $R_Z$ rotations on quantum states would introduce imaginary terms, so the QAML model adopts $R_Y$ rotation to prepare the initial mapped states, where classical samples determine the rotation angles of qubits. Let $x_i^j$ denote the $j$th element of the sample $x_i$ scaling to the range $[-1,1]$, and its corresponding qubit encoding is
\begin{equation}
\label{E22}
\begin{aligned}
\begin{array}{ccc}
|\varphi(x_i^j)\rangle=\cos(\frac{\pi}{2}x_i^j)|0\rangle+\sin(\frac{\pi}{2}x_i^j)|1\rangle.
\end{array}
\end{aligned}
\end{equation}
Then, the qubit encoding of $x_i$ corresponds to the tensor product state
\begin{equation}
\label{E3}
\begin{aligned}
\begin{array}{ccc}
|\varphi_{i}\rangle=|\varphi(x_i^1)\rangle \otimes |\varphi(x_i^2)\rangle \otimes...\otimes |\varphi(x_i^N)\rangle.
\end{array}
\end{aligned}
\end{equation}

In the QAML model, the parameterized quantum circuit is responsible for transforming the Hilbert space of samples. The variable parameters are continuously optimized in iterations to obtain the optimal Hilbert space for separating samples of different classes. Parameterized quantum circuit, also called ansatz, generally adopts a multi-layer circuit structure, where each layer contains a series of unitary operations depending on variable parameters. Ansatz can embed samples into the Hilbert space that classical metric learning methods cannot represent. Hardware-efficient ansatz, one of the common ansatzes, has strong expressivity with fewer layers\cite{zoufal2019quantum}, and it is widely applied in Noisy Intermediate-Scale Quantum (NISQ)devices. Hardware-efficient ansatz adopts a layered circuit layout\cite{kandala2017hardware}, where each layer consists of interleaved 2-qubits unitary modules. Let $W^k_{ij}(\theta)$ denote the unitary module acting on the neighboring qubit pair $(i, j)$ in the $k$th layer. The unitary operation in the $k$th layer can be written as
\begin{equation}
\label{E3}
\begin{aligned}
\begin{array}{ccc}
W^k(\theta)=\prod_{i\in N_1}{W_{i,(i+1)}^k}(\theta)\prod_{j\in N_2}{W_{j,(j+1)}^k(\theta)},
\end{array}
\end{aligned}
\end{equation}
where $N_1$ and $N_2$ represent the odd and even subsets of $[0, N-1]$. For $l_1$-layer structure, the ansatz can be written as $W(\theta)=\prod_{k=1}^{l_1}W^k(\theta)$.

The dimension of the mapping quantum state is exponential in the input dimension. As the input dimension increases, the dimension of the mapping quantum states will be much larger than the input dimension. In some machine learning tasks, the QAML model may be expected to have a smaller output dimension to facilitate subsequent subroutine execution, the QAML model needs to add some unitary models to adjust the output dimension. A primary strategy is to add dimension reduction operation following the repeated encoding layer $U_1(x_i)$ to reduce the output dimension\cite{cong2019quantum}. The dimension reduction operation is shown in Fig.3 (b). Firstly, alternating 2-qubit unitary modules act on two neighboring qubits to entangle the mapping features. Then, one qubit of each module is measured, and the measurement result is used to control the unitary operation acting on another qubit. Let $Q_{ij}^{k}=tr_i(P^k_{ij})$ denote the operation acting on the $(i,j)$ qubit pair in the $k$th layer, where $tr_{i}$ represents the partial operation on the $i$th qubit. $P_{ij}^k=|0\rangle\langle0|\otimes P^0_{ij}+ |1\rangle\langle1|\otimes P^1_{ij}$ is the controlled unitary, which represents to perform single-qubit unitary $P^0_{ij}$ or $P^1_{ij}$ on the second register according to the measurement result of the first qubit, then $Q^{k}=\prod_{i,j}Q_{ij}^{k}$ represent the dimension reduction operation of the $k$th layer. Assume the dimension reduction operation includes $l_2$ layers, and the output state can be reduced to $2^{N/(2^{l_2})}$-dimensional Hilbert space.

Classical metric learning based on triplet loss function needs three identical CNN to extract the features of the triplet set $(x_i^a, x_i^p, x_i^n)$. To reduce the requirement of parameterized quantum circuits, the QAML model encodes the triplet set on two-qubit basis, then interferes with positive and negative sample pairs by a Hadamard gate. The inner products for the positive and negative sample pair are got in parallel by measuring the expectation of $\sigma_z$ observables with respect to 2 qubits of basis state. Let $|\varphi^a_{i}\rangle$, $|\varphi^p_{i}\rangle$, and $|\varphi^n_{i}\rangle$ represent the states of anchor sample $x^a_i$, positive sample $x^p_i$, and negative sample $x^n_i$, respectively. Firstly, the QAML model prepares a superposition state
\begin{equation}
\label{E22}
\begin{aligned}
\begin{array}{ccc}
|\varphi_{i}\rangle=\frac{1}{{2}}|\varphi_i^{a}\rangle_{s}|0\rangle_1|0\rangle_2+\frac{1}{{2}}|\varphi_i^{a}\rangle_{s} |1\rangle_1|0\rangle_2+\frac{1}{2}|\varphi_i^{n}\rangle_{s}|0\rangle_1|1\rangle_2
                    +\frac{1}{2}|\varphi_i^{p}\rangle_{s}|1\rangle_1|1\rangle_2
\end{array}
\end{aligned}
\end{equation}
for the triplet set $(x^a_i,x^p_i,x^n_i)$, where $s$ is sample register, and 1 and 2 denote ancilla registers for basis states. Metric learning based on triplet loss function requires a specific margin between the samples of different classes. To construct the margin, we replace
$|\varphi_i^{n}\rangle_{s}|0\rangle_1|1\rangle_2$ with
\begin{equation}
\label{E23}
\begin{aligned}
\begin{array}{ccc}
|\varphi_i^{n}\rangle_{s}|0\rangle_1(\frac{\alpha}{\sqrt{\alpha^2+1}}|0\rangle_2+\frac{1}{\sqrt{\alpha^2+1}}|1\rangle_2)
\end{array}
\end{aligned}
\end{equation}
and $|\varphi_i^{p}\rangle_{s} |1\rangle_1|1\rangle_2$ with
\begin{equation}
\label{E24}
\begin{aligned}
\begin{array}{ccc}
|\varphi_i^{p}\rangle_{s}|1\rangle_1(-\frac{\alpha}{\sqrt{\alpha^2+1}}|0\rangle_2+\frac{1}{\sqrt{\alpha^2+1}}|1\rangle_2),
\end{array}
\end{aligned}
\end{equation}

where $\alpha$ is the parameter determining the margin. $|\varphi_{i}^a\rangle$, $|\varphi_{i}^p\rangle$, and $|\varphi_{i}^n\rangle$ may not be in the optimal Hilbert space for separating samples of different classes. Then, the parameterized quantum circuit $W(\theta)_s\otimes I_1 \otimes I_2$ acts on $|\varphi_i\rangle$, where $I_1 $ and $I_2$ denote the identity operations acting on ancilla registers 1 and 2, and $W(\theta)_s$ represents the ansatz acting on the sample register $s$. The system gets the state
\begin{equation}
\label{E25}
\begin{aligned}
\begin{array}{ccc}
|\varphi_{i}^{'}\rangle=\frac{\sqrt{2\alpha^2+1}}{2\sqrt{\alpha^2+1}}|\varphi_i^{00}\rangle_{s}|0\rangle_1|0\rangle_2+\frac{\sqrt{2\alpha^2+1}}{2\sqrt{\alpha^2+1}}|\varphi_i^{10}\rangle_{s}|1\rangle_1|0\rangle_2\\
\\

+\frac{1}{2\sqrt{\alpha^2+1}}|\varphi_i^{01}\rangle_{s}|0\rangle_1|1\rangle_2
                    +\frac{1}{2\sqrt{\alpha^2+1}}|\varphi_i^{11}\rangle_{s}|1\rangle_1|1\rangle_2,
\end{array}
\end{aligned}
\end{equation}
where $|\varphi_i^{00}\rangle_{s}=W(\theta)_s(\frac{\sqrt{\alpha^2+1}}{\sqrt{2\alpha^2+1}}|\varphi_i^{a}\rangle_{s}+ \frac{\alpha}{\sqrt{2\alpha^2+1}}|\varphi_i^{n}\rangle_{s})$, $|\varphi_i^{10}\rangle_{s}=W(\theta)_s(\frac{\sqrt{\alpha^2+1}}{\sqrt{2\alpha^2+1}}|\varphi_i^{a}\rangle_{s}-\frac{\alpha}{\sqrt{2\alpha^2+1}}|\varphi_i^{p}\rangle_{s})$,
$|\varphi_i^{01}\rangle_{s}=W(\theta)_s|\varphi_i^{n}\rangle_{s}$, $|\varphi_i^{11}\rangle_{s}=W(\theta)_s|\varphi_i^{p}\rangle_{s}$.

As $W(\theta)_sW^\dagger(\theta)_s=I$, the inner product acting on the state pairs $|\varphi_i^{00}\rangle$ and $|\varphi_i^{01}\rangle$ or $|\varphi_i^{10}\rangle$ and $|\varphi_i^{11}\rangle$ will counteract the effect of $W(\theta)$ and $W^\dagger(\theta)$. An effective strategy is to perform the repeated encoding operation $U_1(x_i)$ on $|\varphi_{i}^{'}\rangle$, which not only solves the problem of the unitary operation and its conjugate operation counteracting effects of each other in the inner product calculation process but also extends the addressable Hilbert space. After the repeated encoding operation $U_1(x_i)$, the system gets the state
\begin{equation}
\label{E361}
\begin{aligned}
\begin{array}{ccc}
|\varphi_{i}^{'}\rangle=\frac{\sqrt{2\alpha^2+1}}{2\sqrt{\alpha^2+1}}|\varphi_i^{00'}\rangle_{s}|0\rangle_1|0\rangle_2+\frac{\sqrt{2\alpha^2+1}}{2\sqrt{\alpha^2+1}}|\varphi_i^{10'}\rangle_{s}|1\rangle_1|0\rangle_2\\
\\
+\frac{1}{2\sqrt{\alpha^2+1}}|\varphi_i^{01'}\rangle_{s}|0\rangle_1|1\rangle_2
                    +\frac{1}{2\sqrt{\alpha^2+1}}|\varphi_i^{11'}\rangle_{s}|1\rangle_1|1\rangle_2,
\end{array}
\end{aligned}
\end{equation}
where $|\varphi_i^{00'}\rangle_{s}= U_1(x_i)|\varphi_i^{00}\rangle_{s}$, $|\varphi_i^{10'}\rangle_{s}= U_1(x_i)|\varphi_i^{10}\rangle_{s}$, $|\varphi_i^{01'}\rangle_{s}= U_1(x_i)|\varphi_i^{01}\rangle_{s}$ and $|\varphi_i^{11'}\rangle_{s}= U_1(x_i)|\varphi_i^{11}\rangle_{s}$.

\subsection{Triplet loss function}
A simple method of computing inner products between sample pairs is the Hadamard classifier method\cite{blank2020quantum}. In this method, two samples are firstly projected into orthogonal subspaces, spanned by standard basis states of one ancilla register. Then, a Hadamard gate acts on the standard basis states to interfere with two samples in the 2-dimensional subspaces. Finally, the inner product between two samples is got by measuring the expectation value of $\sigma_z$ for the ancilla register. The triplet loss function, consisting of inner products for positive and negative sample pairs, needs to compute the weighted sum of inner products for sample pairs, where the weight of positive sample pairs is $+1$, and the weight of negative sample pairs is $-1$. The states of the triplet sets have been prepared on the two-qubit standard basis, shown in Equ.\ref{E361}. The QAML model consists of two ancilla registers, Ancilla register 2 is used to build the inner products of sample pairs. The QAML model adopts one Hadamard gate acting on ancilla register 2 to interfere with sample pairs. If only the expectation of the observable $\sigma_z$ for the ancilla register 2 is measured, the QAML model will get the sum of the inner products for positive and negative sample pairs. The QAML model adds another register (Ancilla register 1) to distinguish between different sample pairs, and measuring the expectation with respect to the $\sigma_z$ operator can get the weights of sample pairs. So the QAML model not only measures the expectation of the observable $\sigma_z$ with respect to ancilla registers 1 but also the expectation for ancilla registers 2. The expectation on two ancilla registers is

\begin{equation}
\label{E37}
\begin{aligned}
\begin{array}{ccc}
\langle \sigma^1_z,\sigma_z^2\rangle
=\frac{\sqrt{2\alpha^2+1}}{4\sqrt{\alpha^2+1}}\langle\varphi_i^{00'}|\varphi_i^{01'}\rangle-\frac{\sqrt{2\alpha^2+1}}{4\sqrt{\alpha^2+1}}\langle\varphi_i^{10'}|\varphi_i^{11'}\rangle\\
\\
=\frac{1}{{4\sqrt{\alpha^2+1}}}(\langle \varphi_i^{n} |W^{\dagger}(\theta)U_1^{\dagger}(x_i^n)U_1(x_i^a)W(\theta)|\varphi_i^{a}\rangle\\
\\
-\langle\varphi_i^{p}| W^{\dagger}(\theta)U_1^{\dagger}(x_i^p)U_1(x_i^a)W(\theta)|\varphi_i^{a}\rangle-\frac{\alpha}{\sqrt{\alpha^2+1}}),\\
\end{array}
\end{aligned}
\end{equation}
where $\frac{\alpha}{\sqrt{\alpha^2+1}}$ represents the margin for separating positive and negative samples.
With the help of classical computation, one gets the triplet loss function
\begin{equation}
\label{E38}
\begin{aligned}
\begin{array}{ccc}
L_l(\theta, |\varphi_i^a\rangle, |\varphi_i^p\rangle, |\varphi_i^n\rangle)=[0, 4\sqrt{\alpha^2+1}\langle\sigma_z^1, \sigma_z^2\rangle]_+.
\end{array}
\end{aligned}
\end{equation}

In practical applications, one batch of samples may contain multiple triplet sets, so the QAML model needs to add a index register to distinguish different triplet sets.
Let one batch of samples include $m$ triple sets. $|\varphi_i^{a}\rangle_s$, $|\varphi_i^{p}\rangle_s$ and $|\varphi_i^{n}\rangle_s$ of Equ.\ref{E22} are replaced by the superposition states $|\widetilde{\varphi}_i^{a}\rangle_{s,d}=\frac{1}{\sqrt{m}}\Sigma_{j=im}^{(i+1)m-1}|\varphi_j^{a}\rangle_{s}|j\rangle_d$, $|\widetilde{\varphi}_i^{p'}\rangle_{s,d}=\frac{1}{\sqrt{m}}\Sigma_{j=im}^{(i+1)m-1}|\varphi^p_{j}\rangle_{s}|j\rangle_d$, and $|\widetilde{\varphi}_i^{n'}\rangle_{s,d}=\frac{1}{\sqrt{m}}\Sigma_{j=im}^{(i+1)m-1}|\varphi^n_{j}\rangle_{s}|j\rangle_d$ to construct the loss function for this batch, where the subscript $d$ denotes the index register. The QAML model performs Equ\ref{E361}-\ref{E38} and yields the expectation value of the observable $\sigma_z$ with respect to ancilla register 1 and 2 as
\begin{equation}
\label{E3}
\begin{aligned}
\begin{array}{ccc}
\langle \sigma^1_z, \sigma^2_z\rangle=-\frac{1}{4m(\sqrt{\alpha^2+1})}\sum_{i=1}^{m}(\langle \varphi_i^{n'}| W^{\dagger}(\theta)U_1^{\dagger}(x_i^n)U_1(x_i^a)W(\theta)|\varphi_i^{a'}\rangle\\
\\
-\langle \varphi_i^{p'}|W^{\dagger}(\theta)U_1^{\dagger}(x_i^p)U_1(x_i^a)W(\theta)|\varphi_i^{a'}\rangle-\frac{\alpha}{\sqrt{\alpha^2+1}}),
\end{array}
\end{aligned}
\end{equation}
which corresponds to the weighted sum of the inner products for one batch samples.

\subsection{Adversarial samples generation}

Metric learning is vulnerable to adversarial attacks. Attackers usually adopt adding small and imperceptible perturbations on natural samples to generate adversarial samples for deceiving metric learning models. Adversarial attacks make metric learning models unable to accurately distinguish positive and negative samples and give rise to misclassification. Miyato\cite{miyato2018virtual} proposed an adversarial training method, where ambiguous but critical adversarial samples are generated based on the gradients of natural samples and added to the training set\cite{liu2020vulnerability}. This method effectively fights against white-box attacks and improves the robustness of the model. Inspired by this method, we developed a quantum adversarial samples generation method. Considering the efficiency of the triplet loss function, we do not create adversarial samples corresponding to all natural samples. Anchor samples in the triplet loss function are used twice to compute the inner products of positive and negative sample pairs. The adversarial samples corresponding to anchor samples can provide more valuable information for adversarial training, so the QAML model only build adversarial samples corresponding to anchor samples.

Let $|\varphi_a^*\rangle$ denote the adversarial sample corresponding to the anchor sample $|\varphi_a\rangle$. According to the characteristics of adversarial attacks, $|\varphi_a^*\rangle$ is far from the positive sample $|\varphi_p\rangle$ but close to the negative sample $|\varphi_n^*\rangle$, and this characteristic makes the QAML model hard to build accurate metric distances. According to Ref\cite{kurakin2016adversarial}, adversarial attacks generated along the direction of gradient ascent will produce the strongest disturbance to metric learning, so we develop a quantum gradient ascent method to generate adversarial samples.
Let $\bigtriangledown_i^a=({(\bigtriangledown_i^a)}^1,{(\bigtriangledown_i^a)}^2,...,{(\bigtriangledown_i^a)}^N)$ denote the gradient vector of the loss function $L_l(\theta, |\varphi_i^a\rangle, |\varphi_i^p\rangle, |\varphi_i^n\rangle)$ with respect to the anchor sample $|\varphi_i^a\rangle$, where the element $(\bigtriangledown_i^a)^j= \partial(L_l(\theta, |\varphi_i^a\rangle, |\varphi_i^p\rangle, |\varphi_i^n\rangle))/\partial(|\varphi_i^a\rangle^j)$ is the partial derivation of the loss function with respect to the $j$th element of $|\varphi_i^a\rangle$.

The QAML model may encounter many attacks. One of the common attacks is the white-box attack, under which the attackers have complete information about the QAML model, including the loss function implemented by parameterized quantum circuit, so that they can compute the gradients of the loss function with respect to gate parameters. Let the QAML model suffer from the functional adversarial attack\cite{mcclean2017hybrid} (one kind of white-box attacks), under which each element of quantum states is influenced by the attack independently. According to the idea of gradient ascent, the adversarial anchor sample $|\varphi^{i*}_a\rangle$ can be written as
\begin{equation}
\label{E4}
\begin{aligned}
\begin{array}{ccc}
|\varphi^{a*}_i\rangle=\frac{1}{\sqrt{1+\lambda^2||\bigtriangledown_i^a||_2^2}}(|\varphi^a_i\rangle+ \lambda\bigtriangledown_i^a|\varphi_i^a\rangle),
\end{array}
\end{aligned}
\end{equation}
where $\lambda=(\lambda_1,\lambda_2,...,\lambda_N)$ is a constant vector used to control the disturbance within a specified bound. Usually, $\lambda$ is determined by the problem to be solved and its upper bound is $||\lambda||_p\leq \varepsilon$, where $||\cdot||_p$ denotes $l_p$-norm.

Let $V(\lambda\nabla_i^a)=v(\lambda_1{(\nabla_i^a)}^1)\otimes...\otimes v(\lambda_N{(\nabla_i^a)}^N)$ denote the unitary acting on the anchor sample $|\varphi_i^a\rangle$ to generate the adversarial sample $|\varphi_i^{a*}\rangle$, where $v(\lambda_j{(\nabla_i^a)}^j)$ represents the unitary operation acting on the $j$th element of $|\varphi_i^a\rangle$. It is expected that $v(\lambda_j{(\nabla_i^a)}^j)$ has small impact on the state $|\varphi_i^a\rangle$, so $V(\lambda\nabla_i^a)$ is close to the identity operator $I$. $v(\lambda_j{(\nabla_i^a)}^j)$ can be implemented by the rotation operation
\begin{equation}
\label{E3}
\begin{aligned}
\begin{array}{ccc}

R_y(2\beta)=
\begin{bmatrix}
\cos(\beta), &-\sin(\beta)\\
\\
\sin(\beta), &\ \ \ \cos(\beta)\\
\end{bmatrix},

\end{array}
\end{aligned}
\end{equation}
where $\beta=\arccos(1+\lambda_j{(\nabla_i^a)}^j)$. As the QAML model only adopts anchor samples to generate adversarial samples, we define the unitary operation to generate adversarial sample as
 \begin{equation}
\label{E3}
\begin{aligned}
\begin{array}{ccc}
V^{'}(\lambda\nabla_i^a)=V(\lambda\nabla_i^a)_{s}\bigotimes I_1\bigotimes \prod_2^0+I_{s}\bigotimes I_1 \bigotimes\prod_2^1,
\end{array}
\end{aligned}
\end{equation}
where $V(\lambda\nabla^a_i)$ acts on the sample register $s$ only when the ancilla register 2 is $|0\rangle$, and $I_{s}$ and $I_{1}$ mean the identity unitary $I$ acting on registers $s$ and 1, respectively. Fig.3 (c) shows the schematic of generating adversarial samples, where $U^{'}_1(x_i^a)=V^{'}(\lambda\nabla_i^a)U_1(x_i^a)$ replaces $U_1(x_i^a)$ to generate the adversarial sample $|\varphi_i^{a*}\rangle$. In the QAML training process, the parameters $\theta$ are optimized by alternatively minimizing the loss function $L_l(\theta, |x_i^{a}\rangle, |x_i^p\rangle, |x_i^n\rangle)$ and $L_l(\theta, |x_i^{a*}\rangle, |x_i^p\rangle, |x_i^n\rangle)$, where natural and adversarial samples are respectively served as input.

The core work of generating adversarial samples is to compute the partial deviation ${(\nabla_i^a)}^j$. Many methods can be adopted to calculate ${(\nabla_i^a)}^j$, such as the finite difference scheme and parameter shift rule\cite{crooks2019gradients,schuld2019evaluating,mitarai2018quantum}. The parameter shift rule has faster convergence in the training process, making it more suitable for NISQ devices. $(\nabla_i^a)^j$ is evaluated using the parameter shift rule
\begin{equation}
\label{E41}
\begin{aligned}
\begin{array}{ccc}
\partial(L_l(\theta, |x_i^a\rangle, |x_i^p\rangle, |x_i^n\rangle)/\partial((x_i^a)^j)
\\
\\=\frac{1}{2}(L_l(\theta, |x_{i,j}^{a+}\rangle, |x_i^p\rangle, |x_i^n\rangle)-L_l(\theta, |x_{i,j}^{a-}\rangle, |x_i^p\rangle, |x_i^n\rangle)),
\end{array}
\end{aligned}
\end{equation}
where $x_{i,j}^{a\pm}=x_{i}^a\pm\frac{\pi}{2}e^j$, and $e^j$ is the unit vector with only the $j$th qubit being 1. According to Equ\ref{E41}, one partial derivative can be got by evaluating the loss function twice.

\section{Numerical simulations and discussions}

In this section, we adopt the PennyLane software framework\cite{bergholm2018pennylane} to demonstrate the performances of the QAML model. The QAML model is implemented by a hybrid quantum-classical algorithm, where the quantum device and classical optimizer cooperate to implement parameter optimization. RMSProp\cite{mukkamala2017variants} optimizer serves as a classical optimizer with a learning rate of 0.01. Our first work is to demonstrate the performance of the QAML model on the MNIST dataset, consisting of $28\times28$-dimensional grayscale images of handwritten digits $0\sim9$. The QAML model focuses on binary classification tasks, so only two categories of handwritten digits, '$0$' and '$1$', are chosen to form data sets. As NISQ devices have limited circuit depth and qubits, the QAML model first reduces samples into 2-dimensional vectors using the principal component analysis (PCA) method. The training and test sets contain 100 samples, respectively, where 50 samples are from class '$0$' and 50 samples come from class '$1$'.

Fig.4 shows the distributions of test samples in the Hilbert space. Simulation results show that samples from different classes are pushed apart with a larger margin and become linearly separable after performing the QAML model. Fig.5 (colorbar figure) shows the inner products between test sample pairs (the larger the inner product, the smaller the distance). The QAML model without adding adversarial samples can be viewed as the general quantum metric learning model, named as the QML model. Panel (a) shows the inner products of sample pairs before performing the QML or QAML models. Panel (b) shows the inner products of sample pairs after performing the QML model, where the training set only includes natural samples. Panel (c) denotes the inner products of sample pairs after completing the QAML model, where the training set consists of natural and adversarial samples. Before training, the inner products for sample pairs of the same and different classes have little difference. This phenomenon means that samples from different categories are close to each other and are difficult to separate. After performing the QML model, the inner products for negative sample pairs become smaller (close to 0), indicating that the distances between samples from different categories begin to get larger. After performing the QAML model, the inner products for negative sample pairs are going to -1, smaller than the values obtained through the QML model. This result indicates that the distance between samples of different categories after executing the QAML model is greater than that after executing the QML model. Let $d_{i}$ represent the average inner product of all sample pairs from the same class, shown in Table.1. $d_{o}$ denotes the average inner product of all sample pairs from different classes, offered in Table.2. The result shows that the average inner product $d_{o}$ in the case of adding adversarial samples is smaller than that without adversarial samples, regardless of training or test sets. This result also means that the QAML model can obtain a larger separation margin than the QML model. We also can find that the average inner product $d_{o}$ for test and training samples have little differences, indicating that the QAML model has a good generalization for the unseen test data.

\begin{figure}[H]
	\centering
	\subfigbottomskip=2pt
	\subfigcapskip=-5pt
	\subfigure[Before QAML]{
		\includegraphics[width=0.48\linewidth]{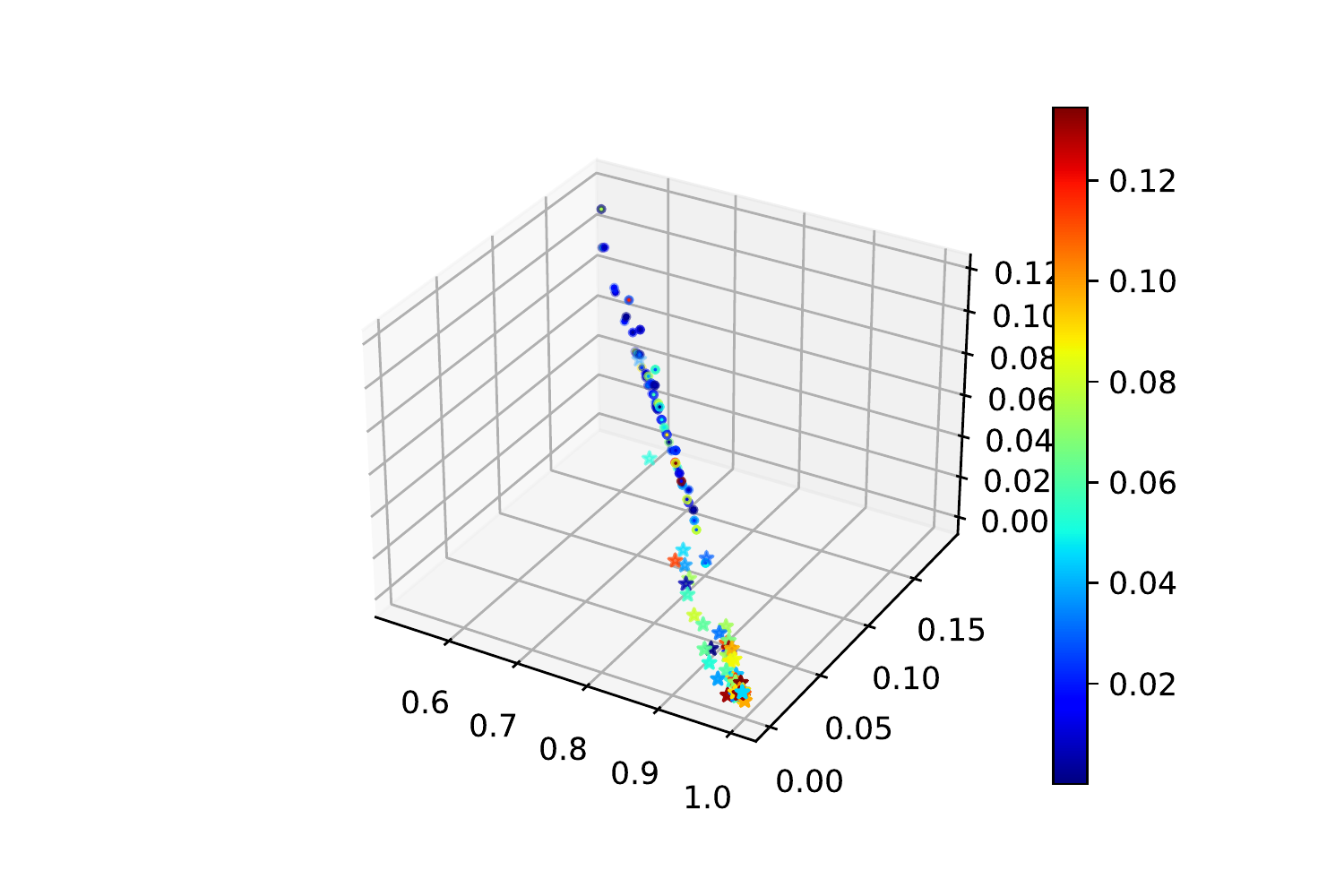}}
	\subfigure[After QAML]{
		\includegraphics[width=0.48\linewidth]{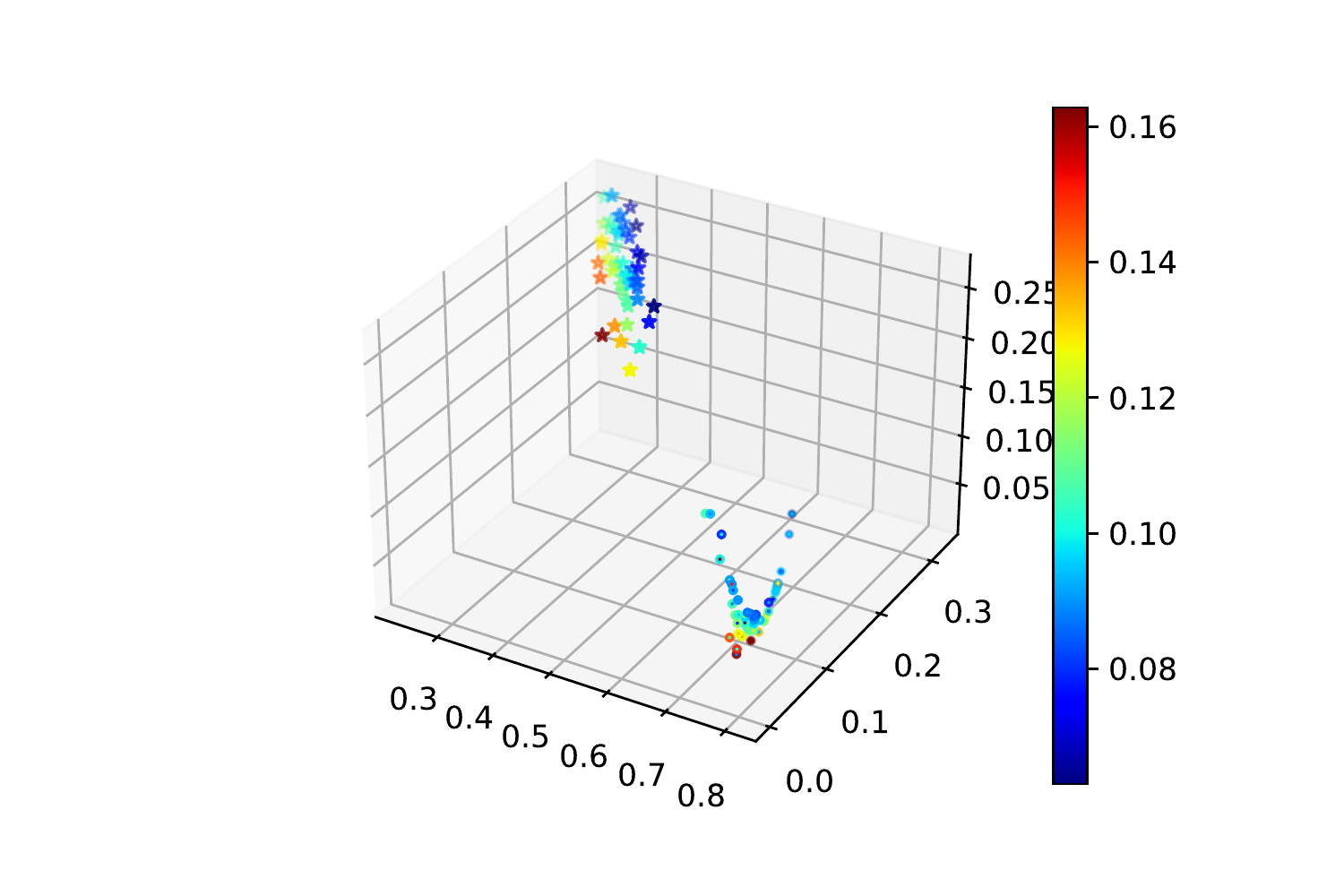}}
	\caption{The distributions of samples in the Hilbert space. 'star' denotes samples from class '0', and 'circle' represent samples from class '1'. Panel (a) shows the distribution of samples before performing the QAML model. Panel (b) shows the distribution of samples after completing the QAML model.}
\end{figure}

\begin{figure}[H]
	\centering  
    \vspace{-1.5cm}
	\subfigbottomskip=2pt 
	\subfigcapskip=-5pt 
	\subfigure[Before QML (QAML)]{
		\includegraphics[width=0.32\linewidth]{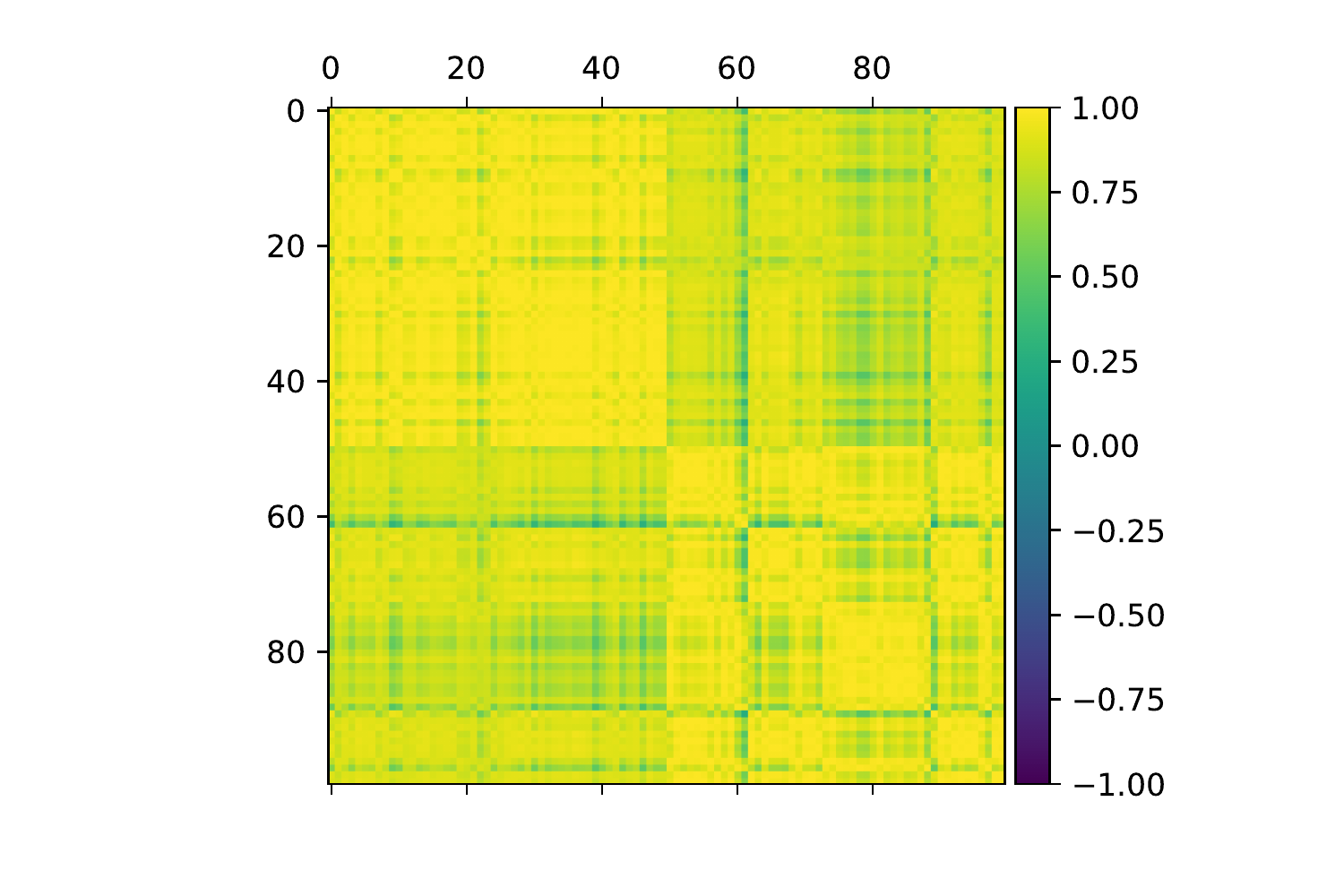}}
	\subfigure[After QML]{
		\includegraphics[width=0.32\linewidth]{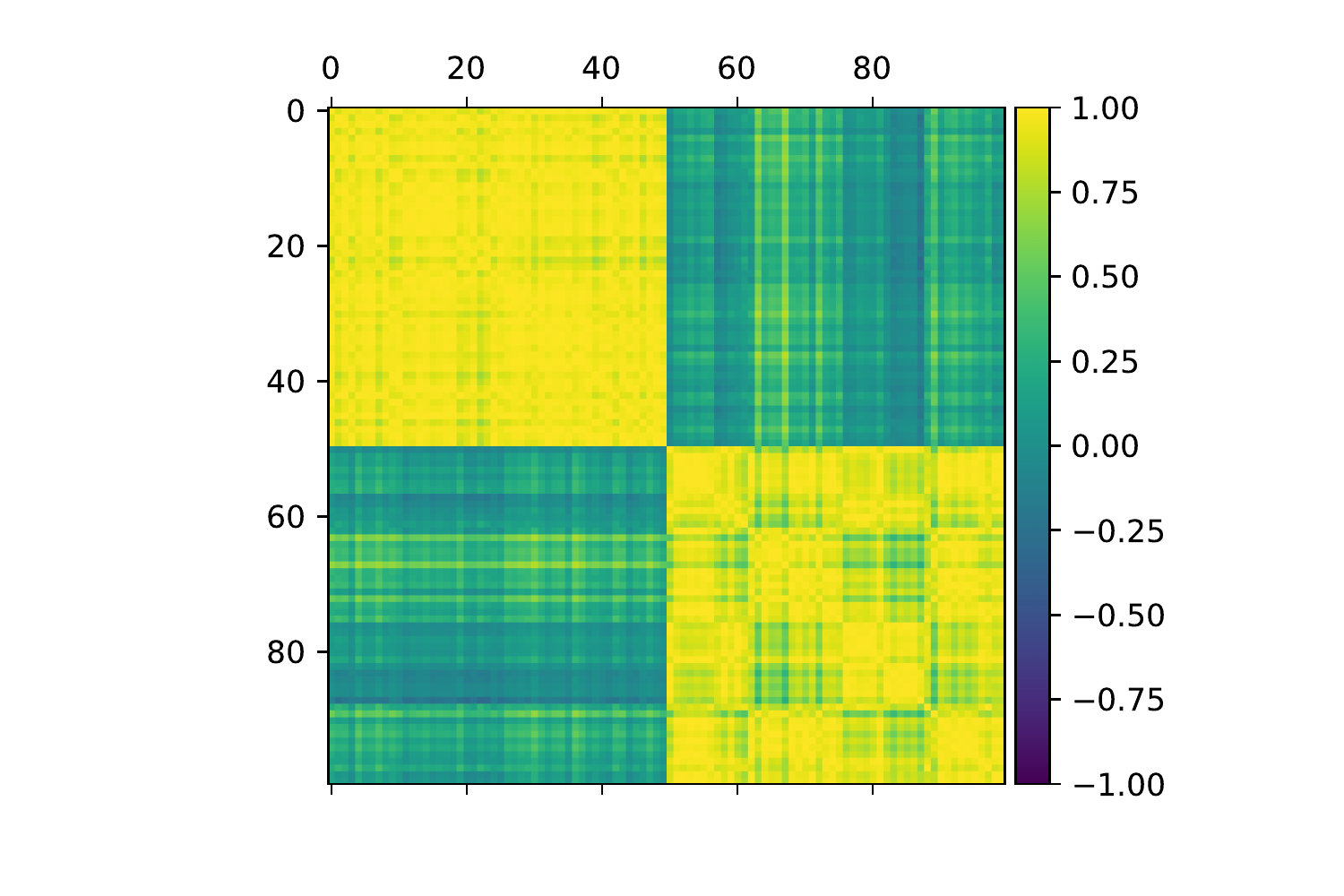}}
	\subfigure[After QAML]{
		\includegraphics[width=0.32\linewidth]{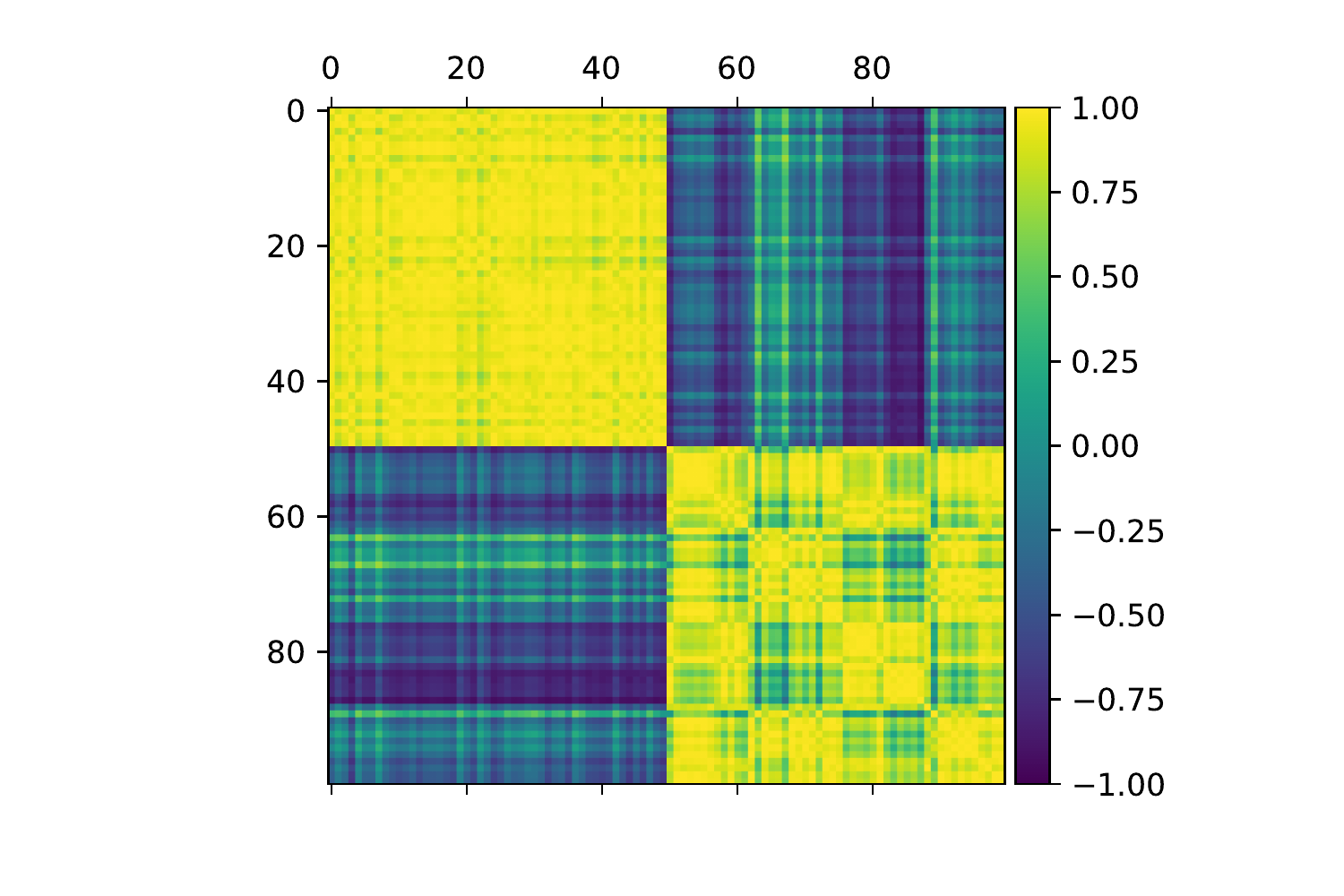}}
	\caption{The inner products between test sample pairs of MNIST. The horizontal and vertical axes represent the indexes of samples. Indexes 0-49 denote the samples from class '0', and indexes 50-99 represent the samples from class '1'. Panel (a) shows the inner products between all sample pairs before performing the QML model, also corresponding to the inner products before performing the QAML model. Panel (b) shows the inner products of test sample pairs, where the QML model is trained through 1000 training epochs but adversarial samples are not added to the training set. Panel (c) shows the inner products of test sample pairs after 1000 training epochs, where adversarial samples are added to the training set.}
\end{figure}

\linespread{1.2}
\begin{table}[H]
\caption{The average inner products $d_{i}$ of sample pairs from the same class (MNIST dataset). The first row describes the average inner products for sample pairs before training, and the second row depicts the inner products for sample pairs after training. The first two columns represent the average inner products for training and test sample pairs, respectively, where adversarial samples are not added to the training set. The last column represent the average inner products for training and test sample pairs, respectively, where adversarial samples are added to the training set.}
\label{tab:1}       
\begin{tabular}{p{55pt}p{55pt}p{55pt}p{55pt}p{55pt}}
\hline\noalign{\smallskip}
Samples      & Training             & Test               & Training+adv      &Test+adv   \\
\noalign{\smallskip}\hline\noalign{\smallskip}
Before       & 0.8280               & 0.8168             & 0.8280            & 0.8168  \\
After        & 0.8348               & 0.8021             & 0.8537            & 0.8249  \\

\noalign{\smallskip}\hline
\end{tabular}
\vspace{-0.5cm}
\end{table}

\linespread{1.2}
\begin{table}[H]
\caption{The average inner products $d_{o}$ of sample pairs from different classes (MNIST dataset). The description of rows and columns is the same as Table 1.}
\label{tab:1}       
\begin{tabular}{p{55pt}p{55pt}p{55pt}p{55pt}p{55pt}}
\hline\noalign{\smallskip}
Samples      & Training            & Test          & Training+adv     &Test+adv   \\
\noalign{\smallskip}\hline\noalign{\smallskip}
Before       & 0.3040              & 0.4787         & 0.3040             & 0.4787  \\
After        &  -0.7971            &  -0.6968       &-0.8326             &  -0.7696  \\

\noalign{\smallskip}\hline
\end{tabular}
\vspace{-0.5cm}
\end{table}

To further verify the separation effects for other data sets, we simulate the performances of the QML and QAML models on Iris dataset. Iris dataset contains 150 samples with 4-dimensional features, where samples $0\sim 49$  belong to class 1, samples $50\sim 99$ belong to class 2, and samples $100\sim 149$ belong to class 3. Samples from classes 2 and 3 are difficult to separate by simple linear functions, so we select them to build a binary data set, where 30 samples of each category are used to construct the training set, and the other 20 samples are served as the test set. Fig.6 shows the average inner products of test sample pairs for Iris dataset. Panels (a), (b), and (c) show the inner products for test sample pairs before performing the QML or QAML model, after performing the QML model, and after performing the QAML model, respectively. Simulation results show that the QAML model also has good separation effects on Iris dataset, superior to the QML model.
Tables 3 and 4 show the average inner products $d_{i}$ and $d_{o}$ for Iris dataset, respectively. Simulation results show that all $d_{i}$ have similar values, indicating that the sample from the same class has relatively stable distances regardless of whether performing the QAML model. Before performing the QML or QAML model, $d_{o}$ has a larger value, which means that samples from different classes are close to each other and are difficult to separate. After performing the QML and QAML models, the average inner products $d_{o}$ get smaller values, where $d_{o}$ of the QAML model has smaller values than that of the QML model. We can find that the QAML model yields a better separation effect than the QML model, and the conclusion is consistent with that got based on MNIST dataset.

\begin{figure}[H]
	\centering  
	\subfigbottomskip=2pt 
	\subfigcapskip=-5pt 
	\subfigure[Before QML (QAML)]{
		\includegraphics[width=0.32\linewidth]{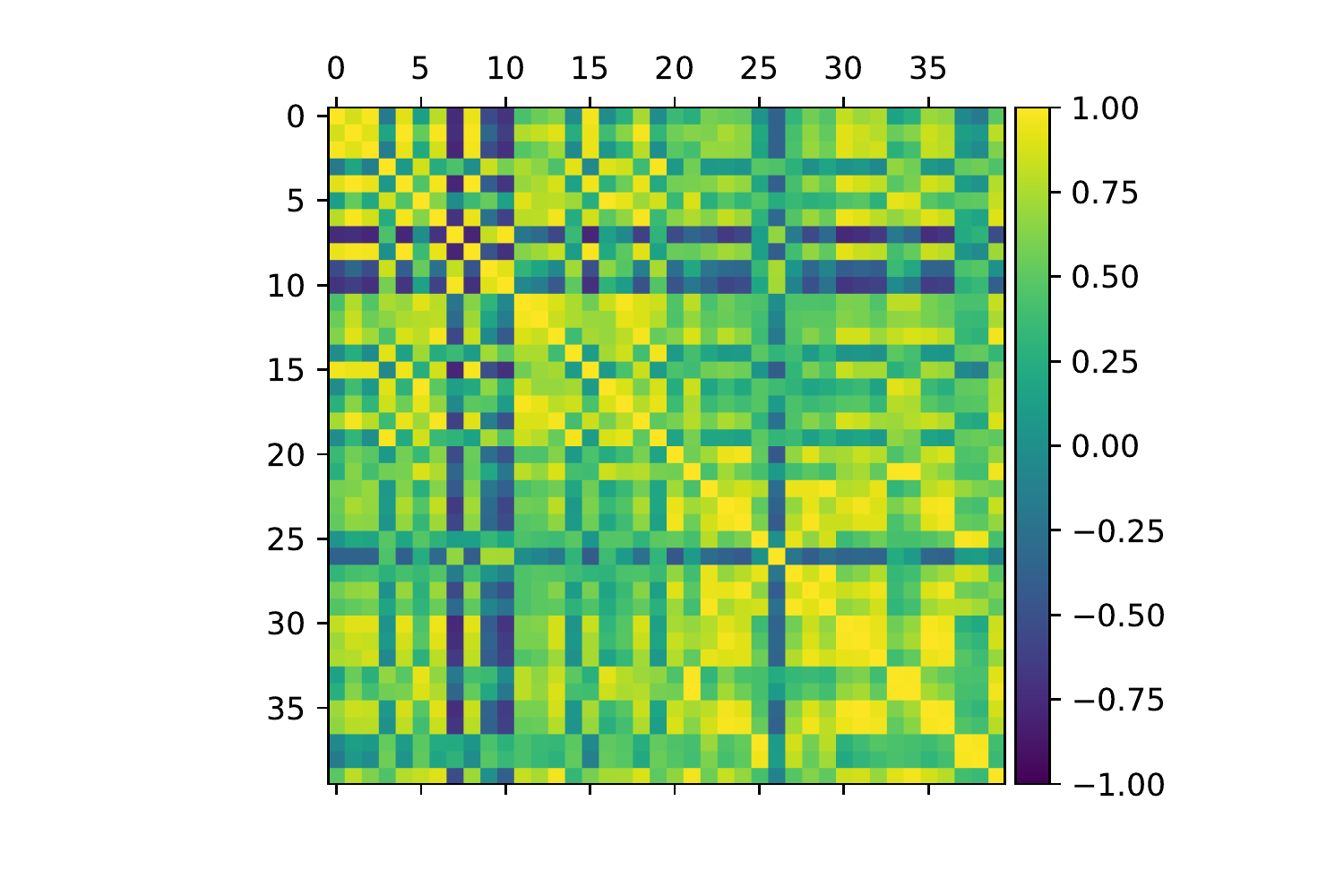}}
	\subfigure[After QML]{
		\includegraphics[width=0.32\linewidth]{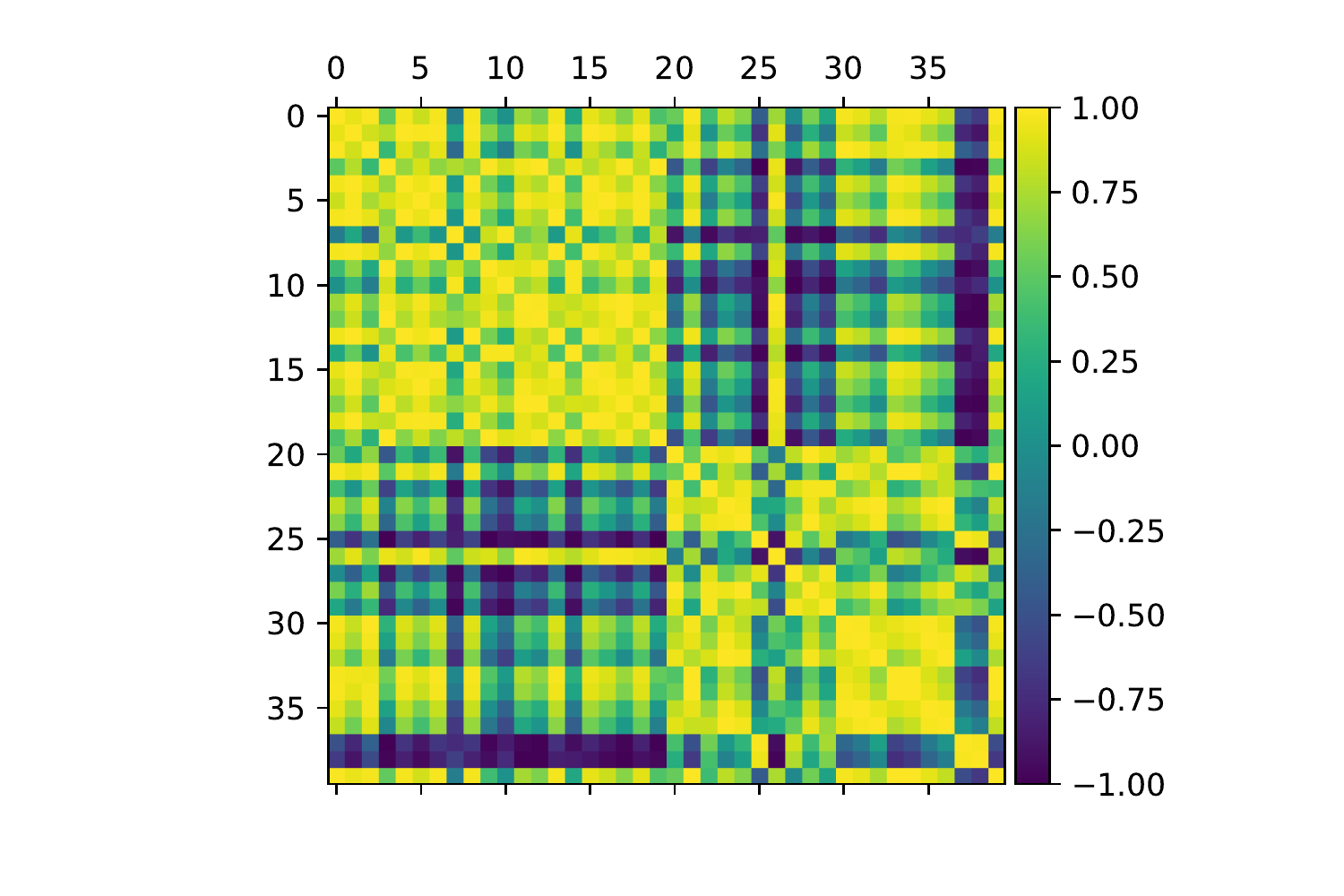}}
	\subfigure[After QAML]{
		\includegraphics[width=0.32\linewidth]{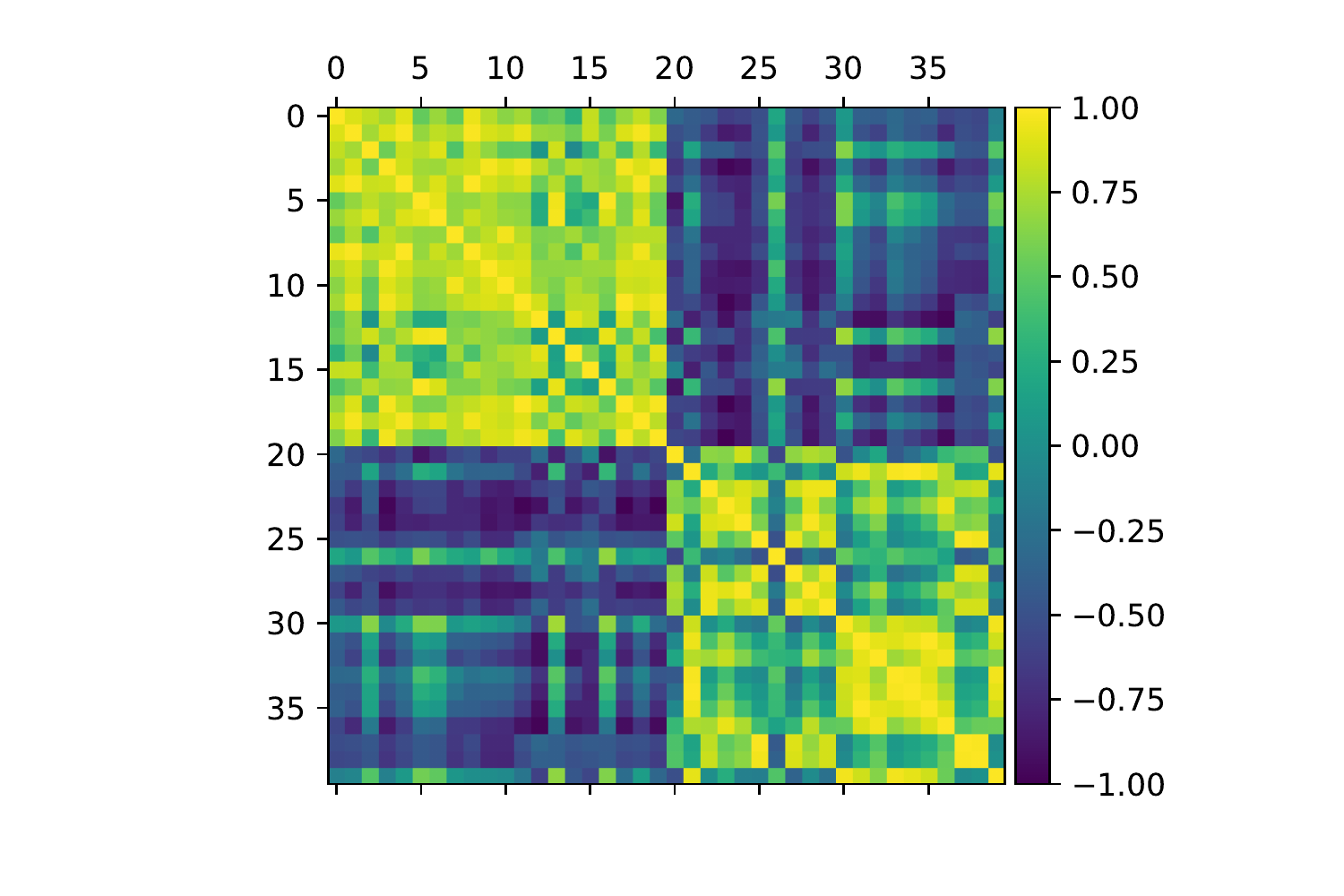}}
	\caption{The inner products for sample pairs of Iris dataset. Indexes 0-19 denote test samples from class 2, and indexes 20-39 represent test samples from class 3. Panel (a) shows the inner products of test sample pairs before performing the QML (QAML) model. Panel (b) shows the inner product of test sample pairs after performing the QML model. Panel (c) shows the inner products of test sample pairs after performing the QAML model.}
\end{figure}
\linespread{1.2}
\begin{table}[H]
\caption{The average inner products $d_{i}$ of samples from the same class (Iris dataset). The description of rows and columns is the same as Table 1.}
\label{tab:1}       
\begin{tabular}{p{55pt}p{55pt}p{55pt}p{55pt}p{55pt}}
\hline\noalign{\smallskip}
Samples      & Training             & Test               & Training+adv      &Test+adv   \\
\noalign{\smallskip}\hline\noalign{\smallskip}
Before       & 0.5065              & 0.5909              & 0.5065              & 0.5909  \\
After        & 0.5473              & 0.6109              & 0.5549               & 0.6544  \\

\noalign{\smallskip}\hline
\end{tabular}
\vspace{-0.5cm}
\end{table}

\linespread{1.2}
\begin{table}[H]
\caption{The average inner products $d_{o}$ of samples from different classes (Iris dataset). The description of rows and columns is the same as Table 1.}
\label{tab:1}       
\begin{tabular}{p{55pt}p{55pt}p{55pt}p{55pt}p{55pt}}
\hline\noalign{\smallskip}
Samples      & Training            & Test          & Training+adv     &Test+adv   \\
\noalign{\smallskip}\hline\noalign{\smallskip}
Before       & 0.3377              & 0.4787          & 0.3377              & 0.4787  \\
After        &  -0.6314            &  -0.3424        & -0.6752             &  -0.4653  \\

\noalign{\smallskip}\hline
\end{tabular}
\vspace{-0.5cm}
\end{table}

Furthermore, we prove the robustness of the QAML model based on the $\epsilon$-robust accuracy proposed in Ref.\cite{guan2020robustness}.
Given a test sample set $\mathcal{S}$ and a smaller threshold $\epsilon$. Let $\rho\in\mathcal {S}$ represent the quantum state of a test sample of $\mathcal{S}$. If $\rho$ and another state $\sigma$ belong to different classes and the inner product between them is larger than the threshold $\epsilon$, then $\sigma$ is viewed as the adversarial sample of $\rho$. If $\rho$ has no adversarial samples within $\epsilon$, $\rho$ is $\epsilon$-robust state. Let $\mu_\epsilon$ denote the $\epsilon$-robust accuracy of $\mathcal{S}$, which is equal to the proportion of $\epsilon$-robust states of the sample set $\mathcal{S}$. Let the threshold be $\epsilon=0.02$. The $\epsilon$-robust accuracies of the QML and QAML models in MNIST dataset are $92\%$ and $100\%$, respectively. The $\epsilon$-robust accuracies of the QML and QAML models on Iris dataset are $91\%$ and $95\%$, respectively. Compared with the QML model, the QAML model improves the robustness by adding the adversarial samples to the training set.
\bibliographystyle{spphys}       
\bibliography{manuscript}   
\end{document}